\documentclass[9pt,twocolumn,twoside]{pnas-new}
\usepackage{natbib}

\templatetype{pnasresearcharticle} 

\title{Understanding Stellar Contamination in Exoplanet Transmission Spectra as an Essential Step in Small Planet Characterization}

\author[a,b,1]{D\'aniel Apai}
\author[a,b]{Benjamin V. Rackham} 
\author[b,c]{Mark S. Giampapa}
\author[d,e]{Daniel Angerhausen}
\author[f]{Johanna Teske}
\author[g]{Joanna Barstow}
\author[h]{Ludmila Carone}
\author[i]{Heather Cegla}
\author[j]{Shawn D. Domagal-Goldman}
\author[h]{N\'estor Espinoza}
\author[i]{Helen Giles}
\author[k]{Michael Gully-Santiago}
\author[l]{Raphaelle Haywood}
\author[m]{Renyu Hu}
\author[n]{Andres Jordan}
\author[l]{Laura Kreidberg}
\author[o]{Michael Line}
\author[p]{Joe Llama}
\author[l]{Mercedes L\'opez-Morales}
\author[q]{Mark S. Marley}
\author[r]{Julien de Wit}

\affil[a]{Steward Observatory and Lunar and Planetary Laboratory, The University of Arizona, Tucson, AZ, USA}
\affil[b]{EOS Team, NASA Nexus for Exoplanet System Science, USA}
\affil[c]{National Solar Observatory, Tucson, AZ, USA}
\affil[d]{Center for Space and Habitability, University of Bern, Switzerland}
\affil[e]{Blue Marble Space Institute of Science}
\affil[f]{Hubble Postdoctoral Fellow, Department of Terrestrial Magnetism, Carnegie Institution of Washington, USA}
\affil[g]{RAS Research Fellow, Astrophysics Group, Department of Physics and Astronomy, UCL, UK}
\affil[h]{Max Planck Institute for Astronomy, Heidelberg, Germany}
\affil[i]{Geneva Observatory, University of Geneva, Switzerland}
\affil[j]{Planetary Environments Laboratory, NASA Goddard Space Flight Center, Greenbelt, MD, USA}
\affil[k]{Kepler/K2 Guest Observer Office NASA Ames Research Center, Moffett Field, CA, USA}
\affil[l]{Harvard–Smithsonian Center for Astrophysics, Cambridge, MA, USA}
\affil[m]{Reny Hu, Jet Propulsion Laboratory, CA, USA}
\affil[n]{Pontific University, Santiago de Chile, Chile}
\affil[o]{Arizona State University, Phoenix, AZ, USA}
\affil[p]{Lowell Observatory, Flagstaff, AZ, USA}
\affil[q]{NASA Ames Research Center, CA, USA}
\affil[r]{Earth and Planetary Sciences Department, MIT}

\leadauthor{Apai} 

\correspondingauthor{\textsuperscript{1}To whom correspondence should be addressed. E-mail: apai@arizona.edu}

\begin{abstract}
Transmission spectroscopy during planetary transits is expected to be a major source of information on the atmospheres of small (approximately Earth-sized) exoplanets in the next two decades. This technique, however, is intrinsically affected by stellar spectral contamination caused by the fact that stellar photo- and chromospheres are not perfectly homogeneous. Such stellar contamination will often reach or exceed the signal introduced by the planetary spectral features. Finding effective methods to correct stellar contamination  -- or at least to quantify its possible range -- for the most important exoplanets is a necessary step for our understanding of exoplanet atmospheres. This will require significantly deepening our understanding of stellar heterogeneity, which is currently limited by the available data.
\end{abstract}

\dates{This manuscript was compiled on \today}
\doi{http://eos-nexus.org/whitepapers/}

\begin{document}

\verticaladjustment{-2pt}

\maketitle
\thispagestyle{firststyle}
\ifthenelse{\boolean{shortarticle}}{\ifthenelse{\boolean{singlecolumn}}{\abscontentformatted}{\abscontent}}{}


\pagebreak

\clearpage


\begin{SCfigure*}
\centering
\includegraphics[width=12.5cm]{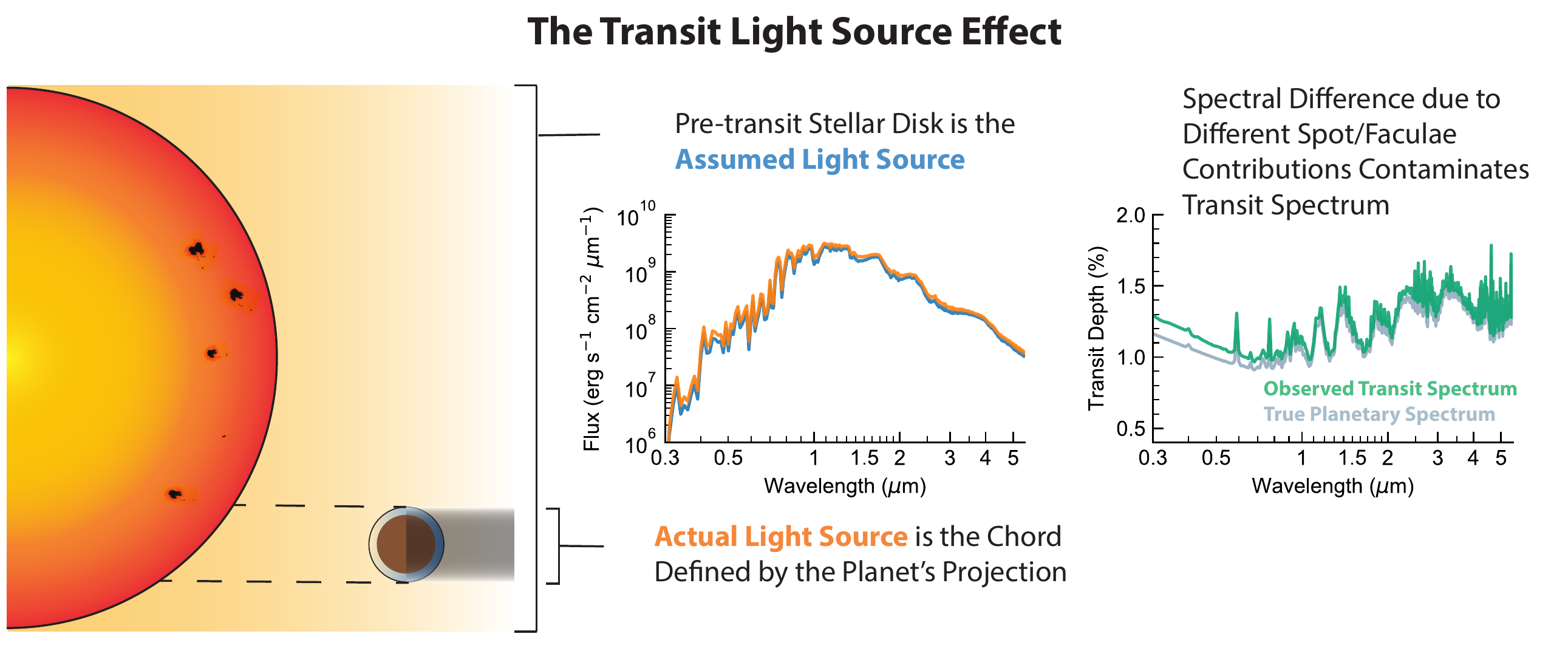}
\caption{In transiting exoplanet transmission spectroscopy the light source is the chord defined by the planet's projection, but its spectrum is not directly observable. The spectral difference between the disk-integrated and chord-integrated stellar spectra will contaminate the transmission spectrum -- the contamination level can often reach or exceed the level of intrinsic planetary features. From \cite{Rackham2018}.}
\label{fig:TLSE}
\end{SCfigure*}

\dropcap{O}ver the past sixteen years our understanding of extrasolar planet atmospheres has been revolutionized by transit spectroscopy. High-precision measurements of the wavelength-dependence of the apparent sizes of transiting exoplanets reveal opacity variations and, through the presence of atmospheric features, probe the presence of key absorbers. Progress and perspectives for exoplanet transmission spectroscopy have been summarized in multiple outstanding reviews \cite{2017arXiv170905941K,2009PASP..121..952D,2015PASP..127..311C,2016PASP..128i4401S, 2016ApJ...817...17G,2017arXiv170507098F} and we only highlight here representative results to illustrate the breadth of knowledge gained from transit spectroscopy. 

Atomic features have been reported in an increasing number of hot jupiters -- examples range from the first pioneering observation of an exoplanet's atmosphere through the Na~{\sc{I}} detection in  HD~209458b \cite{2002ApJ...568..377C} to the more recent K~{\sc{I}} line in XO-2b \cite{2012MNRAS.426.1663S}, to the simultaneous Na~{\sc{I}} and K~{\sc{I}} detections in HD 189733b \cite{2013MNRAS.432.2917P}. 

Molecular features have also been detected, including water in HD~209458b and XO-1b \cite{2013ApJ...774...95D} and TiO in WASP-19b \cite{2017Natur.549..238S}. Furthermore, gradually increasing apparent planet sizes at shorter wavelengths -- interpreted as Rayleigh scattering slopes -- were observed in the hot jupiter HD189733~b \cite[e.g.][]{2007A&A...476.1347P,2013MNRAS.432.2917P,2014ApJ...791...55M, 2015JATIS...1c4002A}. Similar results have been found for the warm exo-neptune GJ 3470b \cite{2015ApJ...814..102D,2017A&A...600A.138C}.

The success of Kepler prime and extended missions, and wide-field ground-based transit surveys have dramatically increased the number of known transiting planets, many of which are suitable for transit transmission spectroscopic characterization with ground-based telescopes, the Hubble Space Telescope ($HST$), and with the upcoming James Webb Space Telescope ($JWST$); this sample will expand through the transiting exoplanet surveys $TESS$ and $PLATO$. In the next two decades this combination of suitable planets and powerful telescopes capable of follow-up observations promises to define our understanding of small exoplanets, including those in the habitable zone \cite{2014SPIE.9143E..20R, 2014ExA....38..249R, 2015ApJ...810...29H, 2015ApJ...809...77S}.

\section*{The Transit Light Source Effect}
{\em The transit light source effect (TLSE) is the contamination of the exoplanet transit spectrum due to the difference between the stellar disk-integrated spectrum and the spectrum of the transit chord} \cite{Rackham2018}, as illustrated in Figure~\ref{fig:TLSE}.  For realistic stars -- i.e., not perfectly homogeneous -- the spectrum of any chord will differ slightly from the disk-integrated spectrum. When obtaining the transmission spectrum of the transiting exoplanet, the light source is the transit chord, the spectrum of which, however, is not directly observable. In lieu of the transit chord's spectrum the pre- and post-transit spectra are measured: these are disk-integrated spectra. 

In most cases the disk-integrated spectra are very similar, but not identical, to the transit light chord's spectrum. Occulted and unocculted starspots, faculae, plages, and flares will introduce slight spectral differences between the disk-integrated and chord-integrated spectra, resulting in the TLSE. This effect is thought to be most important for M stars due to their enhanced stellar activity.

\section*{Current Corrections are Problematic}

The fact that unocculted stellar features impact transmission spectra has been recognized for well over a decade \cite[e.g.,][]{Pont2008}. However, essentially all published stellar contamination corrections have derived the spot (sometimes spot/faculae) areal covering fractions from the photometric variability amplitude (or analogous measurements) for the host stars. Two important and related {\em assumptions} underpin these corrections:
{\em a)} A linear correlation exists between the variability amplitude and the covering fraction of stellar spots (and/or faculae). {\em b)} Most stars have very homogeneous ($>99\%$) photospheres/chromospheres.

\begin{figure}
\centering
\includegraphics[width=1.0\linewidth]{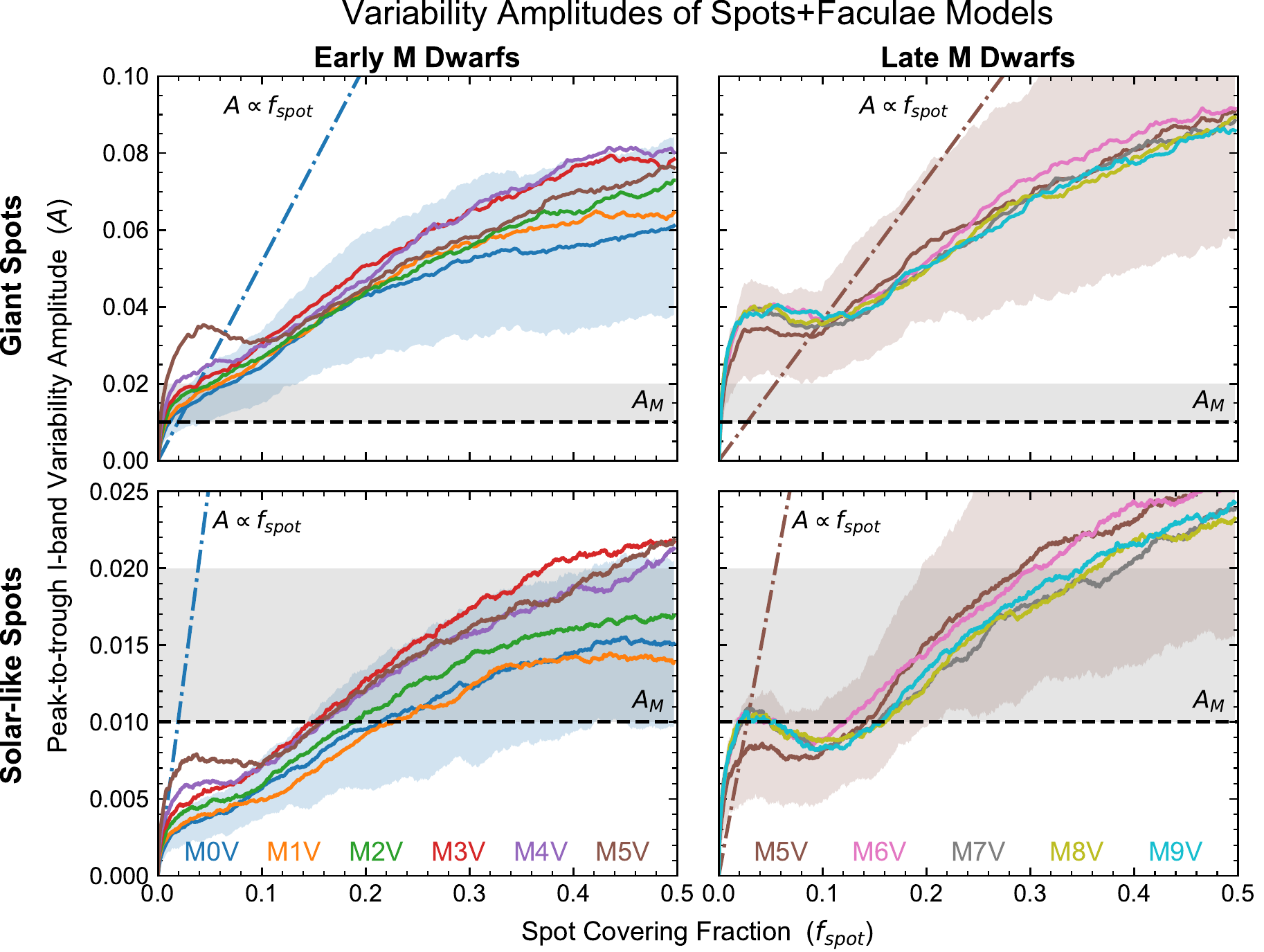}
\caption{The amplitude of photometric variability is often used to estimate the spot covering fraction (an essential parameter to correct for stellar contamination), but multiple realistic simulations show the assumed linear correlation between variability amplitude and spot covering fraction (dot-dashed line) is incorrect. Except for a few exceptional cases (extremely quiet stars), photometric amplitudes cannot be used to reliably determine spot covering fractions. Curves with different colors show the median photometric variability predicted by simulations in which spot coverage is gradually increased. The shaded region shows the 1$\sigma$ scatter of the variability for a given spot covering fraction. Dashed lines show the variability of TRAPPIST-1, while the gray shaded region shows the typical range of photometric variability measured for M-dwarfs by \cite{Newton2016}. Figure from \cite{Rackham2018}.}\label{fig:Amplitudes}
\end{figure}

While reasonable as first-order assumptions, closer inspection reveals that both assumptions are {\em not} correct in general and will often lead to a greatly underestimated stellar spectral contamination (see Figure~\ref{fig:Amplitudes} and \cite{Rackham2018}). 

Assumption (a) is incorrect because photometric variability of an unresolved rotating sphere, in essence, measures the deviation from the rotationally symmetric brightness distribution and {\em not the spot covering} fraction. Independent Monte Carlo models by multiple groups (using somewhat different assumptions and model setups) reached the same conclusion \cite{Jackson2013,Rackham2018}: {\em stellar photometric or spectroscopy variability amplitudes are very insensitive to changes in spot covering fractions}.
(Instead of $\Delta v \propto f_{spot}$, simulations shows that  $\Delta v \propto \sqrt{f_{spot}}$, where $\Delta v$ is variability amplitude and $f_{spot}$ is the spot covering fraction). In contrast, the stellar contamination is directly proportional to the stellar features' surface covering fraction. 

Assumption (b) is problematic because it contradicts a multitude of indirect and direct observational evidence for stellar activity being common across the stellar spectral types and because there is no physical mechanism for main sequence stars that would favor perfectly homogeneous photospheres/chromospheres. 

Unfortunately, essentially all published corrections for unocculted features rely on photometric/spectroscopic variability to constrain spot/facular covering fractions: In the absence of detailed information about the host star's spot properties, the spot properties are generally inferred from the stellar photometric variability, assuming all variability is due to a single spot rotating in and out of view  (e.g., \cite{Pont2008,Desert2011,BertaThompson2011,McCullough2014}. However, this method neglects the likely scenario that there are multiple spots, that spot temperatures may not be uniform, and the possible presence of faculae and plage.

\section*{Solar and Stellar Heterogeneity}

Main-sequence, late-type stars show analogs of atmospheric inhomogeneities that we see in the Sun in the form of spots and faculae.  This heterogeneous structure is a {\em fundamental} property of the solar atmosphere \citep{Golub1980} and, by extension from observational inference, late-type stellar atmospheres as well. These structures are spatially associated with localized concentrations of emergent magnetic fields.  The principal manifestations of quasi-steady magnetic activity relevant to our discussion of stellar contamination of transmission spectra are photospheric faculae, their chromospheric counterparts -- plages -- and spots. 

Faculae and plages are characterized on the Sun by magnetic field strengths less than the photospheric gas equipartition value of $\sim$1,500 G.  Field strengths in sunspot umbrae generally range from  $\sim$ 2,000--4,000 G and with temperatures in the broad range of $\sim$ 50--80\% of the solar effective temperature.  The identification of water vapor lines in sunspot umbral spectra is consistent with the occurrence of low temperatures $\sim$3,000 K \citep{Bernath2002}. The filling factor of the {\em quiet} facular network changes from 15\% at solar minimum to 20\% at maximum \citep[Cycle 21]{Foukal1991}. About half of the 11-year sunspot cycle modulation of disk-integrated Ca {\sc ii} K in the Sun is caused by the change in the fractional area coverage of active regions and half is due to the changing filling factor of the combined quiet and active network elements \citep{Criscuoli2017}.  Therefore, as concluded by \cite{Criscuoli2017}, the 5\% change in cycle-modulated quiet facular area is only about one-half of the total change in this parameter.  Hence, facular filling factors on the Sun roughly range from 15--30\% from solar minimum to solar maximum.  Facular brightenings on the Sun are typically observed on the centerward side of granules with sizes that can extend to about 0.6 arcseconds \citep{Keller2004}. Still today, we have very few direct observations of faculae, including measurements of their effective temperature.
But from the results of simulations of near-surface convection in late-type, main sequence stars \citep{Beeck2015}, we infer facular effective temperatures that can be $\sim$100 K hotter than the photospheric effective temperature in early G dwarfs.  However, the apparent brightness of facular elements and their inferred temperatures can depend on viewing angle, magnetic field strength and wavelength, in addition to spectral type \cite{Norris2017}.

While faculae are present at all latitudes on the solar disk and throughout a solar cycle, sunspots initially emerge at mid-latitudes of $\sim~\pm$30$^{\circ}$.  This mid-latitude activity band would also include transit chords for our solar system viewed as, say, a $Kepler$ Object of Interest (KOI). As the cycle progresses, spots emerge closer and closer to the equator producing the so-called Butterfly Diagram \cite{Maunder1922, Babcock1961}.  Faculae are particularly concentrated near spots where these active regions can be the sources of especially intense chromospheric and coronal emission.  Sunspots have filling factors of $\sim$ 0.1--0.3\% of the visible hemisphere.  However, during solar minimum, numerous days where no sunspots are observed can occur such as happened during the extended minimum of Cycle 23 with over 600 spotless days \citep{Nandy2011}. Grand Minima such as the Maunder minimum are characterized by decades of very few observed sunspots. Observable sunspots that did appear were confined to the southern hemisphere, indicative of parity interactions in the operative solar dynamo \citep{Tobias1997}.

\begin{table}
\caption{Overview of methods used to assess stellar heterogeneity and representative results. \label{Methods}}
\begin{tabular}{lcl}
Method & Sp. Type & Results on Covering Fractions $f$, Refs.\\
\midrule
\multicolumn{3}{l}{Disk-integrated Photometric Variations}  \\
Photometric  & FGK & Lower limits only.   \\
Photometric  & M & $\Delta i$=0.5-4\%,$f_{spot}>1-8\%$ \cite{Jackson2013,Newton2016,Newton2017,Rackham2018}\\
\midrule
Spectropol. & dM1 & $T_{spot}=0.8 T_{phot}$, $f_{spot}>7\%$  \cite{Berdyugina2011}\\ 
\midrule
\multicolumn{3}{l}{Chromospheric Lines} \\
H$\alpha$ absorption & M1-M5 & $f_{facular}>10-26\%$ \citep{Cram1979,Giampapa1985}\\
H$\alpha$ emission & dM & Active regions, kG fields with f $>$50\% \\
\bottomrule
\end{tabular}
\end{table}

It remains poorly understood how the solar photospheric/chromospheric features and their evolution during the solar cycle can be extrapolated to main sequence stars, whose disks remain unresolved. \cite{Rackham2018} presents an up-to-date and detailed overview of the constraints available presently for late-type main sequence stars. We summarized the key points of that discussion in Table~\ref{Methods} and 
will only highlight four high-level conclusions from the body of literature reviewed in \cite{Rackham2018}:

1) Fractional areal coverage of spots and faculae cannot be currently reliably derived for unresolved stars in general. 

2) Photometric variability is very common in main-sequence stars, but it can only provide a {\em lower limit} for spot and facular covering fractions. 

3) For M dwarfs chromospheric line formation may provide additional indirect information on covering fractions. The line formation-based studies argue for very high areal covering fractions ($>$10--50\%), i.e., highly heterogeneous photosphere/chromospheres. For dM stars widely distributed kG-strength fields have been derived with active area covering fractions $f>50\%$.

4) Spectropolarimetric studies of diatomic molecules in dM dwarfs provide an independent measure of spot temperatures which, when combined with photometric variability, provide a lower areal covering limit and argue for significant heterogeneity.

In short, no evidence supports the general assumption that spot covering fractions should in general be very close to 0\%; in contrast, available evidence supports high spot/facular covering fractions for M dwarfs ($\sim$5--50\%).

\section*{Impact on Exoplanet Characterization}
\subsection*{Biased Exoplanet Bulk Densities}

Transit observations provide a direct measurement of the exoplanet radius, which, combined with mass measurements, allows the exoplanet bulk density to be calculated. This parameter provides the first insights into the planetary composition and enables constraints on rocky, icy, and gaseous components \citep[e.g.,][]{howe2014, zeng2016, dorn2017}. However, radius measurements can be biased by the presence of unocculted photospheric heterogeneities, such as spots and faculae \cite{Rackham2018}. For example, if star spots are present in the stellar disk but not occulted by the transiting exoplanet, the observed transit depth will be deeper than the true $(R_{p}/R_{s})^2$ transit depth, leading to an overestimate of the exoplanet's radius. This, in turn, will lead to an underestimate of the exoplanet's density. Errors in radius measurements are amplified by a factor of 3 in density calculations, given the dependence of density on radius ($\rho \propto R^{-3}$), which prompts caution in estimates of volatile contents extrapolated from these measurements.

The effect of stellar contamination on density calculations has been studied in detail for M-dwarf host stars \cite{Rackham2018}, which provide the only feasible option for studying small planet atmospheres in the near-term future. The TRAPPIST-1 system \cite{Gillon2016, Gillon2017, Luger2017} provides an instructive example in this respect. The small stellar radius \cite[$R_{s}=0.1210 \pm 0.0030~R_{\odot}$;][]{Delrez2018} allows the seven roughly Earth-sized transiting planets to produce large transit depths, which enable atmospheric characterization studies with $HST$ \cite{deWit2016, deWit2018} and, in the future, $JWST$ \cite{Barstow2016}. Photospheric modeling, however, suggests that the observed $\sim 1\%$ peak-to-trough variability of the host star \cite{Gillon2016}, typical of field mid-to-late M dwarfs \cite{Newton2016}, is consistent with spot and faculae covering fractions of $f_{\textrm{spot}}=8^{+18}_{-7} \%$ and $f_{\textrm{fac}}=54^{+16}_{-46} \%$, respectively \cite{Rackham2018}. These spot coverages can cause the bulk densities of the TRAPPIST-1 planets inferred from $Spitzer$ 4.5~$\mu$m radius measurements to be underestimated by $\Delta(\rho) = -3^{+3}_{-8} \%$, thus leading to overestimates of their volatile contents \cite{Rackham2018}. The problem would be even more severe if the densities were estimated from I-band transit depths -- as are for many transiting planets -- which would lead to a bias of $-8^{+7}_{-20}\%$. The large possible faculae covering fractions, by contrast, increase the likelihood that faculae may be distributed throughout the stellar disk, both within and without the transit chord, which would lessen their impact on transit observations. As with most stars, however, the spatial distribution of active regions on TRAPPIST-1 are unconstrained presently and their effect on density calculations for the TRAPPIST-1 planets remains to be seen.

\subsection*{Spectral Contamination}

Stellar heterogeneity impacts transmission spectra in multiple ways, which depend on both the relative temperature of the inhomogeneity and its location on the projected stellar disk. In the most straightforward case, active regions can be occulted by an exoplanet during a transit. These events can produce observable changes in transit light curves. Brightening events due to star spot crossings are routinely observed in transit observations \cite{Pont2008, Carter2011, Huitson2013, Pont2013}. If uncorrected, star spot crossings effectively decrease estimates of transit depths, which may mask increases in transit depth due to exoplanetary atmospheric features. Conversely, faculae crossings during transit effectively increase transit depths, which can lead to spurious detections of scattering slopes \cite{Oshagh2014}. 

\begin{SCfigure*}
\centering
\includegraphics[width=10cm]{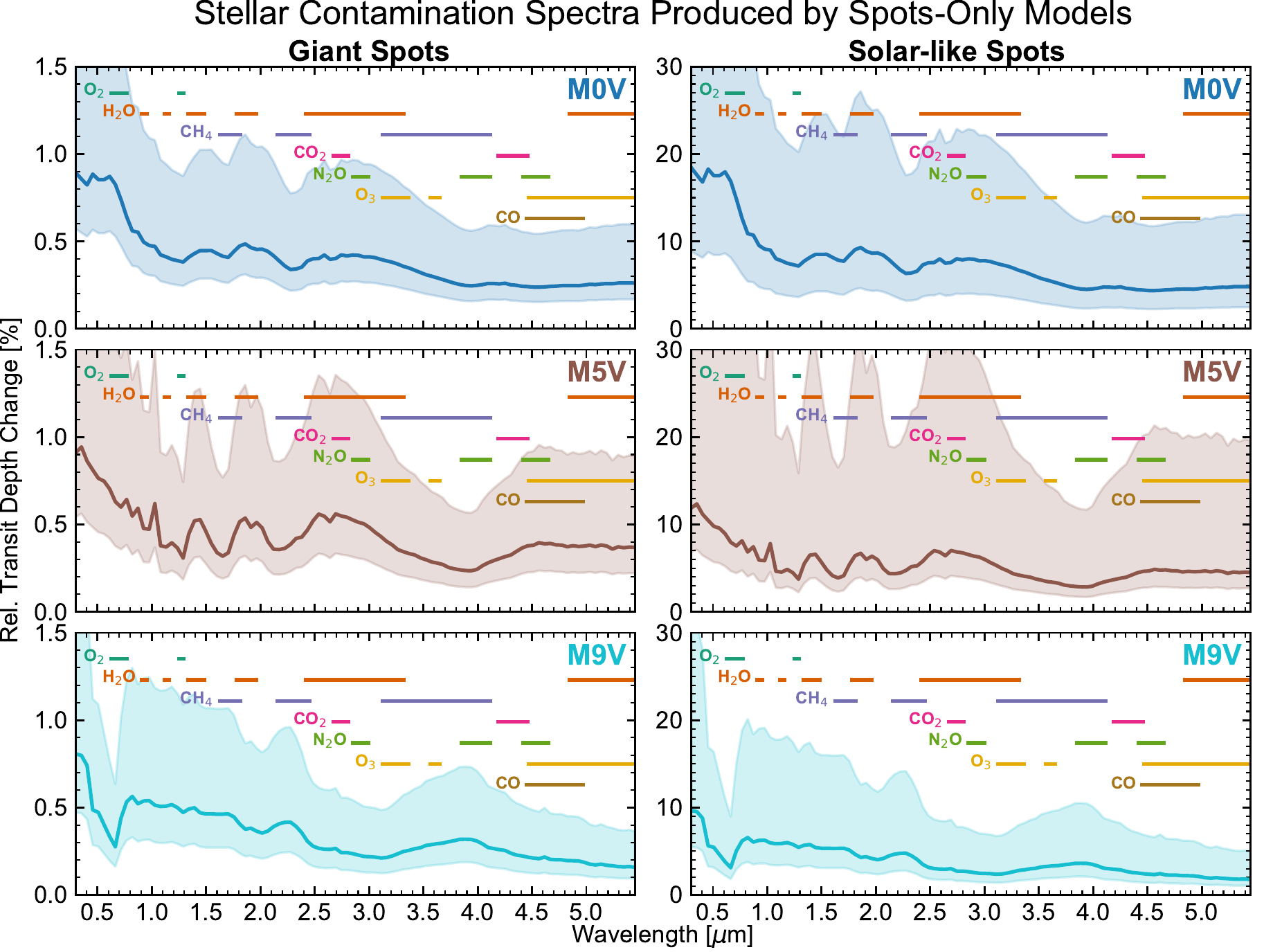}
\caption{Unocculted star spots impart spectral changes in transit depth that can be easily mistaken for absorption or scattering from the exoplanet atmosphere. For a given stellar rotational variability amplitude, a wide uncertainty exists for the level of the stellar contamination present, as illustrated by the shaded regions. For M dwarfs, recent results suggest that stellar contamination signals are roughly 1--15$\times$ the strength of planetary features: for detailed comparison to HST and JWST precision see \cite{Rackham2018}. The scale of the stellar contamination is modulated by many factors, including the spot temperature contrast, typical spot size, and spatial distribution of active regions. Faculae, which have received less study than spots in this context, complicate the issue further, as little is known about their temperature contrasts, covering fractions, and spatial distributions on stars other than the Sun. 
From \cite{Rackham2018}.} \label{fig:contamination_spectra}
\end{SCfigure*}

Active regions located outside of the transit chord also affect transmission spectra. When an exoplanet transits an immaculate photospheric chord but star spots are present in the unocculted stellar disk, the transit chord is effectively brighter per unit area than the disk-averaged brightness. This causes the transit depth to appear deeper than its true $(R_{p}/R{s})^{2}$ value and the exoplanet to appear larger \cite{Pont2013}. For a generalized temperature difference between the transit chord and unocculted spots, the net effect is a chromatic increase in transit depth, strengthening at shorter wavelengths, which can mimic a scattering slope due to $H_{2}$ or aerosols in the exoplanet atmosphere \cite{McCullough2014}. Differences in the opacity of atomic and molecular absorbers between the immaculate photosphere and spotted regions can also impart spectral features on  observed transmission spectra (see Figure~\ref{fig:contamination_spectra}). These features overlap wavelengths of interest for molecular features in exoplanet atmospheres and, in the case of cool M dwarfs, can even be caused by the same molecules of interest (e.g., H$_{2}$O) for the planetary atmosphere being present in the stellar atmosphere as well, which makes disentangling exoplanet atmospheric signals from stellar contamination particularly challenging \cite{Rackham2018}. Simulated observations and retrievals of transit spectra found that M dwarf starspots alone will lead to an overestimate of planetary water vapor by a factos of several \cite{Barstow2015}. To complicate matters further, in contrast to spots, faculae present outside of the transit chord can decrease transit depths, particularly at visual wavelengths \cite{Rackham2017}. This can result in spectral features, such as scattering slopes, being weakened or masked.

The effect of stellar heterogeneity can be particularly problematic for small exoplanets, for which planetary atmospheric signals are small. Combining multiple transit observations, $HST/WFC3$ transmission spectra can reach precisions of 30~ppm per 0.05~$\mu$m wavelength channel \cite{Kreidberg2014}. The same 30~ppm threshold has been suggested as a noise floor for $JWST$ transit observations \cite{Greene2016}. As the strength of the planetary atmospheric transmission feature scales inversely with the square of the stellar radius, these technical considerations constrain observations of terrestrial atmospheric features with these facilities to systems with small host stars, i.e. spectral types later than roughly M5V. However, for these spectral types, stellar contamination signals can be comparable to or up to an order of magnitude larger than planetary atmospheric signals \cite{Rackham2018}. More research is necessary to constrain the degree of stellar heterogeneity for observationally exciting mid-to-late M-dwarf host stars and to disentangle stellar and planetary contributions to exoplanet transmission spectra. (We note that \cite{Deming2017} pointed out an additional, independent effect introduced by the often incorrect treatment of overlapping spectral lines in the planetary and host stars).

\subsection*{The Contamination is Time-varying}

The observed stellar disk's temperature distribution will evolve through at least two processes: stellar rotation and starspot evolution, over the timescale of the faster of the two processes.
Rotational periods for exoplanet host stars range from $<$3.5 days (e.g., TRAPPIST-1) to $>$100 days; starspot evolution timescales are poorly understood for stars different from the Sun, but may occur over timescales of $\sim$10 days (e.g., \cite{Giles2017,Montet2017}). 

The time-evolving stellar contamination poses two challenges: {\em 1)} Combining transit spectra taken at multiple epochs with time differences comparable to or greater than the timescale over which the spectral contamination evolves cannot be simply combined, but have to be first individually corrected for contamination. {\em 2)} In cases when the spectral contamination evolves rapidly (e.g., TRAPPIST-1) the contamination can change significantly even during a single transit observations (typically $\sim$3--5 hours). For a more detailed discussion of the time-variability of the stellar contamination see \cite{Zhang2018}.

\section*{Correction Required}
{ Stellar heterogeneity represents an astrophysical noise source that can contaminate or even overwhelm planetary atmospheric features in transiting exoplanet transmission spectra (Figure~\ref{fig:contamination_spectra}).} {\bf The efficient use of exoplanet transmission spectroscopy requires the development of a robust method for stellar contamination correction or, at least, a method that allows for quantifying the possible levels of contamination.} 

A successful stellar contamination correction requires knowledge of the spectral difference between the disk-integrated light and that of the stellar chord; which, in turn, will probably require a good understanding of the temperature distribution over the stellar disk and in the transit chord. It is possible that the disk-integrated temperature distribution can be determined for certain spectral type stars from temperature-sensitive lines. However, it is likely that for most stars predicting the disk-integrated spectrum will require the following elements: {\em a)} starspot-size distribution function; {\em b)} temperature distribution function for spots and faculae; {\em c)} spot/faculae covering fraction; {\em d)} spot covering fraction at the time of the transit observations.

\section*{Key Questions for Contamination Correction Method}

We identify the following basic questions as essential for developing a robust stellar contamination correction method:

{\em Q1)} How do the starspot and facula properties (size distribution, temperature distribution, spatial distribution) vary with spectral type and stellar activity level?

{\em Q2)} What model components are required to describe the spectral contamination due to stellar heterogeneity?

{\em Q3)} What observations are required by the stellar heterogeneity model to calculate/predict stellar contamination for a given epoch?

Answering these questions will likely require obtaining more constraining stellar observations, developing more realistic photospheric/chromospheric models that can make specific predictions for a range of spectral types and activity levels, and obtaining long-term and/or simultaneous activity indicator observations for each transit event.

\section*{Stellar Heterogeneity and Radial Velocity Modulations}

Time-varying heterogeneous stellar atmospheres can also impact radial velocity (RV)-based planet mass measurements. A thorough review of the different physical and temporal timescales of such features, and current mitigation techniques, can be found in \cite{haywood2015}, but here we stress that temperature variations in a rotating photosphere can be mistaken for planetary signals in RV data or they can influence or mask genuine RV modulations. Understanding how to distinguish stellar RV noise is one of the highest ``tent poles'' holding back the field from finding and measuring the masses of smaller, more Earth-like exoplanets. Thus it is mutually beneficial to both the exoplanet transit and RV communities to better characterize photo- and chromo-spheric heterogeneity.

\section*{Confronting the Problem}
Spot/faculae covering fractions have been systematically underestimated in the exoplanet literature, which has fueled false hopes that stellar contamination is only an issue in exceptional cases. It is important that the stellar contamination is recognized as a challenge, otherwise progress toward understanding its magnitude and correcting for it will be hampered, affecting the science output from HST and JWST. For example, over past cycles multiple proposals that aimed to obtain data to directly test potential transit correction methods for HST and JWST have been declined, leading to the situation where transit data continue to be collected without the small amount of additional data that may be sufficient to correct for the stellar contamination. Without the framework required for testing potential stellar contamination correction methods, otherwise powerful HST and JWST transit spectroscopic datasets will remain uncorrected for stellar contamination.

\section*{Summary}
Transit spectroscopy is a uniquely powerful method to probe the atmospheres of small (sub-jovian) exoplanets. However, spectral contamination is introduced by the spectral difference between the disk-integrated spectrum and the transit chord's spectrum (``transit light source effect'', e.g. \cite{Rackham2018}). This contamination likely represents a critically important challenge to the characterization of small extrasolar planets. The key points of this white paper are as follows:

$\bullet$ Some level of stellar contamination should be expected for most, if not all stars.

$\bullet$ The stellar contamination can impact the spectral slope of the transmission spectrum (commonly used as a proxy of clouds and atmospheric particles) and may also introduce apparent atomic and molecular features

$\bullet$  The amplitude of the contamination is a complex function of the stellar heterogeneity and can range from negligible to levels that may overwhelm intrinsic planetary features

$\bullet$  Correction methods based on stellar variability (photometric or spectroscopic) only probe the non-axisymmetric component of the heterogeneity. These methods possibly/likely underestimate stellar stellar contamination by factors of $\sim$2-10

$\bullet$  Stellar contamination is expected to change on timescales of the stellar rotation and the starspot evolution. For rapidly rotating stars (e.g., TRAPPIST-1) stellar contamination will likely change even during a single transit

$\bullet$ The current understanding of the spatial distribution of the temperature/spectra over stellar disks is insufficient to provide a robust basis for correction methods

\vskip 0.1cm

We recommend the following considerations:

$\bullet$ Formal recognition of the importance of solving the challenges posed by stellar contamination

$\bullet$ Increased interactions between the stellar activity / heliophysics community to develop the physical basis for a contamination correction.

$\bullet$ The definition of the $HST$ and $JWST$ Proposal categories should be adjusted to allow ``calibration proposals'' for astrophysical noise, such as stellar contamination.

\subsection*{References}

\bibliography{pnas-sample}

\end{document}